\documentclass[aps,prl,showpacs,twocolumn,superscriptaddress,letterpaper]{revtex4-1}

\usepackage{graphicx}
\usepackage{dcolumn}
\usepackage{bm}
\usepackage{amsmath, amssymb}%
\usepackage{epstopdf}
\usepackage{color}

\usepackage{hhline}
\usepackage{multirow}

\usepackage{float}
\floatstyle{plaintop}
\restylefloat{table}

\newcommand{\eq}[1]{Eq.~(\ref{#1})}

\newcommand{\rvec}{\ensuremath{\boldsymbol{r}}}
\newcommand{\Rvec}{\ensuremath{\boldsymbol{R}}}

\newcommand{\kvec}{\ensuremath{\boldsymbol{k}}}
\newcommand{\Kvec}{\ensuremath{\boldsymbol{K}}}
\newcommand{\bra}{\langle}
\newcommand{\ket}{\rangle}

\begin{document}


\title{The nuclear contacts and short range correlations in nuclei}

\newcommand*{\HEU }{Racah Institute of Physics, Hebrew University, Jerusalem 91904, Israel}
\newcommand*{\HEUindex}{1}
\affiliation{\HEU}
\newcommand*{\MIT }{Massachusetts Institute of Technology, Cambridge, Massachusetts 02139, USA}
\newcommand*{\MITindex}{2}
\affiliation{\MIT} 
\newcommand*{\TAU }{School of Physics and Astronomy, Tel Aviv University, Tel Aviv 69978, Israel}
\newcommand*{\TAUindex}{3}
\affiliation{\TAU}

  
\author{R. Weiss}
     \affiliation{\HEU}
\author{R. Cruz-Torres}
     \affiliation{\MIT}
\author{N. Barnea}
     \affiliation{\HEU}
\author{E. Piasetzky}
     \affiliation{\TAU}
\author{O. Hen}
     \affiliation{\MIT}

\date{\today}

\begin{abstract}

Atomic nuclei are complex strongly interacting systems and their exact theoretical description is a long-standing challenge. 
An approximate description of nuclei can be achieved by separating its short and long range structure. This separation of scales stands at the heart of the nuclear shell model and effective field theories that describe the long-range structure of the nucleus using a mean-field approximation. We present here an effective description of the complementary short-range structure using contact terms and stylized two-body asymptotic wave functions. The possibility to extract the nuclear contacts from experimental data is presented. Regions in the two-body momentum distribution dominated by high-momentum, close-proximity, nucleon pairs are identified and compared to experimental data. The amount of short-range correlated (SRC) nucleon pairs is determined and compared to measurements. Non-combinatorial isospin symmetry for SRC pairs is identified. The obtained one-body momentum distributions indicate dominance of SRC pairs above the nuclear Fermi-momentum. 

\end{abstract}

\maketitle


The atomic nucleus is one of the most complex systems in nature.
One of the main challenges in describing nuclei is understanding the short interparticle part of the 
nuclear wave function. The challenge stems from the complicated nucleon-nucleon interaction and the large 
density of the nucleus. The latter causes all the relevant scales of the system (nucleon size, average distance, 
and interaction range) to be comparable, making effective theoretical descriptions very demanding. 
On the other hand, detailed understanding of these short-range correlations (SRCs) is important for neutron-star structure and
the nuclear symmetry energy \cite{frankfurt08b, hen15, Cai:2015xga,Hen:2016ysx},
the bound nucleon and free neutron structure functions \cite{weinstein11,hen11,Hen:2013oha,Hen:2016kwk,Hen12,Chen:2016bde},
neutrino-nucleus interactions and neutrino oscillation experiments \cite{Gallagher:2011zza,fields13,Fiorentini13b,acciardi14,Weinstein:2016inx}, and more.

Current mean-field nuclear theories describe well various static properties of nuclei,
 but fail to describe the dynamic effects of SRCs.
Ab-initio many-body calculations 
\cite{wiringa14,Carlson:2014vla, Hagen:2015yea, Rios:2009gb, Rios:2013zqa, Lonardoni:2017egu} are still limited to 
light nuclei and/or to soft interactions that regulate the 
short-range/high-momentum parts of the nuclear wave function. 
Therefore, effective theories are still needed to describe medium and heavy nuclei and to identify the main
physical process at short distances  \cite{ryckebusch15, vanhalst12,colle15,CiofidegliAtti:1995qe}. 

In the last decade there was
a significant progress in describing SRCs in dilute
Fermi systems. It was shown that 
if the interaction range $r_0$ is much shorter than the average interparticle 
distance $d$, and the scattering length $a_s$, a {\it contact} theory can be used to 
describe the system \cite{Tan08a, Tan08b, Tan08c, Braaten12}. 
A series of relations between different observables and 
the probability of finding a particle pair in a close proximity emerge.
The contact theory was studied in great detail theoretically, and validated experimentally,
 for ultra-cold fermi gases \cite{Tan08a, Tan08b, Tan08c, Braaten12,
 gandolfi11, Stewart10, Kuhnle10, partridge05, werner09, schirotzek10, sagi12}.

For nuclei, several experimental observations resemble those of cold atomic systems \cite{hen15b, Weiss14}.
However, in nuclei, the short-range interaction is about $0.5-1.5$ fm,
the average distance between nucleons is about 2.5 fm,
and the scattering length is about -20 fm and 5 fm for
the spin-singlet and spin-triplet channels, respectively.
Therefore, the possibility to generalize the contact
theory to nuclear systems is not obvious.
Nevertheless, a generalized nuclear theory was recently presented which addresses the factorization of the nuclear many-body wave function at short distances~\cite{Weiss:2015mba}. Few of its predictions
were verified, yet more convincing theoretical and experimental
results must be provided to prove
that indeed it is adequate for describing nuclear SRCs.

Many features of nuclear SRCs are well known and should
be properly explained by any candidate theory.
Recent scattering experiments indicate that SRC
 pairs account for 20$\%$ - 25$\%$ of the nucleons in the nucleus and practically all
 nucleons with momentum above the Fermi momentum ($k_F$) 
\cite{frankfurt93,egiyan02,egiyan06,fomin12,tang03,piasetzky06, subedi08, korover14,hen14}.
They are predominantly in the form of neutron-proton ($np$) SRC pairs with large relative momentum ($k>k_F$),
and small center-of-mass (c.m.) momentum  ($K<k_F$). Here, $k_F \sim 255 \rm{MeV/c} = 1.3 \rm{fm}^{-1}$ 
is the typical Fermi momentum of medium and heavy nuclei. These, and results of theoretical studies, indicate that the high-momentum ($k > k_F$) tail of the nuclear momentum distribution is dominated by SRC and described using a factorized wave function for the c.m. and relative momentum distributions of the pairs which results in similar two-body densities for different nuclei \cite{ryckebusch15, vanhalst12, colle15, CiofidegliAtti:1995qe, Alvioli:2016wwp,
 Alvioli:2013qyz, neff15, Frankfurt81, Frankfurt88,Forest:1996kp}. For recent reviews see \cite{Hen:2016kwk, Atti:2015eda}.
Between these well-established properties and 
the generalized contact formalism there is a seemingly unsolved tension,
as the latter's predictions involving two-body
momentum distributions, are only satisfied
for very high momentum, $k> 3 k_F \approx 4$ fm$^{-1}$, and not for 
for lower momentum $k_F < k < 3k_F$.

In this work, we will show that the generalized nuclear contact formalism
can indeed describe SRCs in nuclei also in this lower momentum range. 
A direct agreement with both recent experimental data and with variational Monte Carlo (VMC) calculations will be presented.
We will also discuss the nontrivial manner in which information on 
SRC is encapsulated in the nuclear two-body momentum distributions.
The values of the nuclear contacts for various nuclei will be extracted 
using the VMC two-body distributions
in coordinate and momentum space, separately, and also using experimental data.
We find all three approaches to yield consistent values.
Last, the VMC one-body momentum distributions are compared
to the contact-formalism predictions, confirming the experimental observation
that they are dominated by SRCs for momentum larger than $k_F$.


{\it Generalized contact theory for nuclei --}
The original contact theory was formulated for systems with significant scale separation.
Consequently, the Bethe-Peierls boundary condition can be used, leading 
at short interparticle distance to a factorized asymptotic
wave-function  of the form~\cite{Braaten12}:
\begin{equation}
  \Psi \xrightarrow{r_{ij} \to 0} \varphi(\rvec_{ij}) A_{ij}(\Rvec_{ij}, \{\rvec\}_{k \neq ij}).
\label{eq:awf_1}
\end{equation}
Here $\varphi(\rvec_{ij})$ is an asymptotic two-body wave function,
and $A_{ij}$ is a function of the residual $A-2$ particle system.
The scale separation allows replacing the short-range interaction with a boundary condition, 
and to ignore all partial waves but $s$-wave, leading to
$\varphi(\rvec_{ij})=(1/r_{ij}-1/a_s)$. In momentum space, this factorized wave function 
leads to a high momentum tail, valid 
for $|a_s|^{-1}, d^{-1} \ll k \ll r_0^{-1}$, that is given by: $n(k) \to C/k^4$, where 
  $C = 16 \pi ^2 \sum_{ij}^{ } \bra A_{ij} | A_{ij} \ket$ 
is known as the {\it contact}.

To generalize this formalism to nuclear systems we need to consider two main points:
(1) different partial waves might be significant, and therefore a sum over all possible 
nucleon-nucleon channels $\alpha$ must be introduced,
and (2) as full scale separation does not exist, the asymptotic two-body channel 
wave-functions $\varphi_\alpha$ are taken from
the solution of the nuclear zero-energy two-body problem.
Therefore, the factorized asymptotic wave-function takes the form
\begin{equation}\label{eq:awf_2}
  \Psi \xrightarrow{r_{ij} \to 0} \sum_{\alpha}^{ } \varphi _{\alpha} (\rvec_{ij}) 
  A^{\alpha}_{ij} (\Rvec_{ij}, \{\rvec\}_{k \neq ij}) \;,
\end{equation}
similar to the independent-pair approximation~\cite{Gomes:1957zz},
where the index $ij$ corresponds to $pn$, $pp$, and $nn$ pairs~\cite{Suplementary}.

In this work we will consider only the main channels contributing to SRCs, 
namely, the $pn$ deuteron channel ($\ell=0, 2$ and $s=1$ coupled to $j=1$) 
and the singlet $pp$, $pn$, and $nn$ $s$-wave channel ($\ell=s=j=0$).
Using Eq. (\ref{eq:awf_2}), asymptotic expressions for the one- and two-body momentum densities 
can be derived \cite{Weiss:2015mba}:
\begin{align}\label{eq:1bmdist}
  n_{p}(\kvec) =& 2 C_{pp}^{s=0} |\tilde{\varphi}^{s=0}_{pp} (\kvec)|^2 +
   C_{pn}^{s=0} |\tilde{\varphi}^{s=0}_{pn} (\kvec)|^2 
   \cr & +
   C_{pn}^{s=1} |\tilde{\varphi}^{s=1}_{pn} (\kvec)|^2
\end{align}
\begin{align}\label{eq:2bmd}
F_{pp}(\kvec) =& C_{pp}^{s=0} |\tilde{\varphi}^{s=0}_{pp} (\kvec)|^2
   \cr
F_{pn}(\kvec) =& C_{pn}^{s=0} |\tilde{\varphi}^{s=0}_{pn} (\kvec)|^2 +C_{pn}^{s=1} |\tilde{\varphi}^{s=1}_{pn} (\kvec)|^2
\end{align}
and the same when replacing $n$ with $p$.
Here, $C_{ij}^{\alpha}$ are the nuclear contacts that
determine the number of pairs in a given two-body channel,
$n_N(\kvec)$ is the one-body momentum distribution,
and $F_{NN}(\kvec)$ is the relative two-body momentum distribution. 
$F_{NN}(\kvec)=\int d\Kvec F_{NN}(\kvec,\Kvec)$,
where $F_{NN}(\kvec,\Kvec)$ is the probability of finding a pair of nucleons 
with relative momentum $\kvec $, and center-of-mass (c.m.) momentum $\Kvec$.
Similarly,
$\rho_{NN}(\rvec)$ describes the probability to find a pair of nucleons with relative
distance $\rvec$.
The subscripts N, and NN, stand for the type of nucleon/nucleon-pairs considered.
Clearly, $n_{p(n)}(\kvec) = 2 F_{pp(nn)}(\kvec)+ F_{pn}(\kvec)$ \cite{Weiss:2015mba}. 
Equivalent two-body coordinate space densities for $\rho_{NN}(\rvec)$ are given by replacing 
$\tilde{\varphi}(\kvec)$ with $\varphi(\rvec)$ in Eq.~(\ref{eq:2bmd}), while keeping 
the same nuclear contacts. 
We note that in deriving Eq.~(\ref{eq:1bmdist})
the center-of-mass momentum of the pairs was assumed 
to be much smaller than $k$. 

We choose to normalize $\tilde{\varphi}(\kvec)$  such that 
$\int_{k_{F}}^{\infty}|\tilde{\varphi}(\kvec)|^2 d\kvec = 1$. 
Using this normalization, and Eq.~(\ref{eq:1bmdist}), the fraction of the one-body momentum density above $k_F$ is given by:
\begin{equation}
  \frac{\int_{k_{F}}^{\infty}n(\kvec)d\kvec}{\int_{0}^{\infty}n(\kvec)d\kvec} = 
        \frac{C_{nn}^{s=0}+C_{pp}^{s=0}+C_{pn}^{s=0}+C_{pn}^{s=1}}{A/2} ,
\label{eq:norm}
\end{equation}
where $n(\kvec)=n_n(\kvec)+n_p(\kvec)$, A is the number of nucleons in the nucleus 
and $C_{NN}^{s}/(A/2)$ gives the fraction of the one-body momentum density above the Fermi momentum due to each type of SRC pair.


{\it Ab-initio nuclear two-body densities --}
Recent progress in quantum monte-carlo techniques allows performing ab-initio many-body calculations of 
nuclear structure for nuclei as heavy as $^{12}$C \cite{wiringa14,Carlson:2014vla}. Furthermore,
cluster variational monte-carlo (CVMC) provides a way to obtain nuclear structure calculations for
$^{16}$O and $^{40}$Ca ~\cite{Lonardoni:2017egu}.
These calculations are done using the AV18 and UX potentials, and result in one- 
and two-body nucleon densities in coordinate and momentum space. 

The detailed study of the relation between two-body densities and two-nucleon 
knockout measurements is only now starting \cite{Weiss:2015mba, Alvioli:2016wwp,Alvioli:2013qyz}.

When examining the two-body densities at high relative momentum,
certain care should be taken to separate SRC pairs
from non-correlated pairs with high relative momentum.
Two nucleons that form an SRC pair are close to each other,
each have high momentum, their relative momentum is high, and their c.m. momentum is low.
However, not all nucleon pairs with high relative momentum are necessarily SRC pairs.
For example, a particle with momentum $k_1\approx 4k_F$, and any uncorrelated ``mean-field" particle
at rest $k_2\approx 0$, will yield a pair with high relative momentum $k\approx 2 k_F$, 
and c.m. momentum $K\approx 2 k_F$.
In such cases, the high c.m momentum is a signature for uncorrelated pairs.
As we examine pairs with larger and larger relative momentum, this
scenario becomes less and less probable as the probability of finding
a nucleon with high momentum falls fast with the momentum, i.e. its easier to find
two nucleon with momentum $\approx 2k_F$ then one nucleon with momentum $\approx 4k_F$.

There are two ways to access regions in the two-body momentum distribution dominated by SRC pairs,
with minimal mean-field nucleon contamination. 
One is to integrate over the pairs c.m. momentum but request a very large relative momentum,
which ensures that the pair is truly an SRC pair.
This explains why Ref. ~\cite{Weiss:2015mba} observed scaling between the one and
two-body densities only for momentum much larger than $k_F$.
The alternative approach is to consider pairs with high relative momentum $\kvec$,
and low c.m. momentum $\Kvec$. The cut on $\Kvec$
reduces the contributions from mean field nucleons significantly, and identifies SRC pairs
with lower relative momentum. 
It should be noted that in the limit of heavy nuclei the contribution of uncorrelated nucleon pairs with low c.m. momentum could increase.

\begin{figure} [b]
\includegraphics[width=7cm]{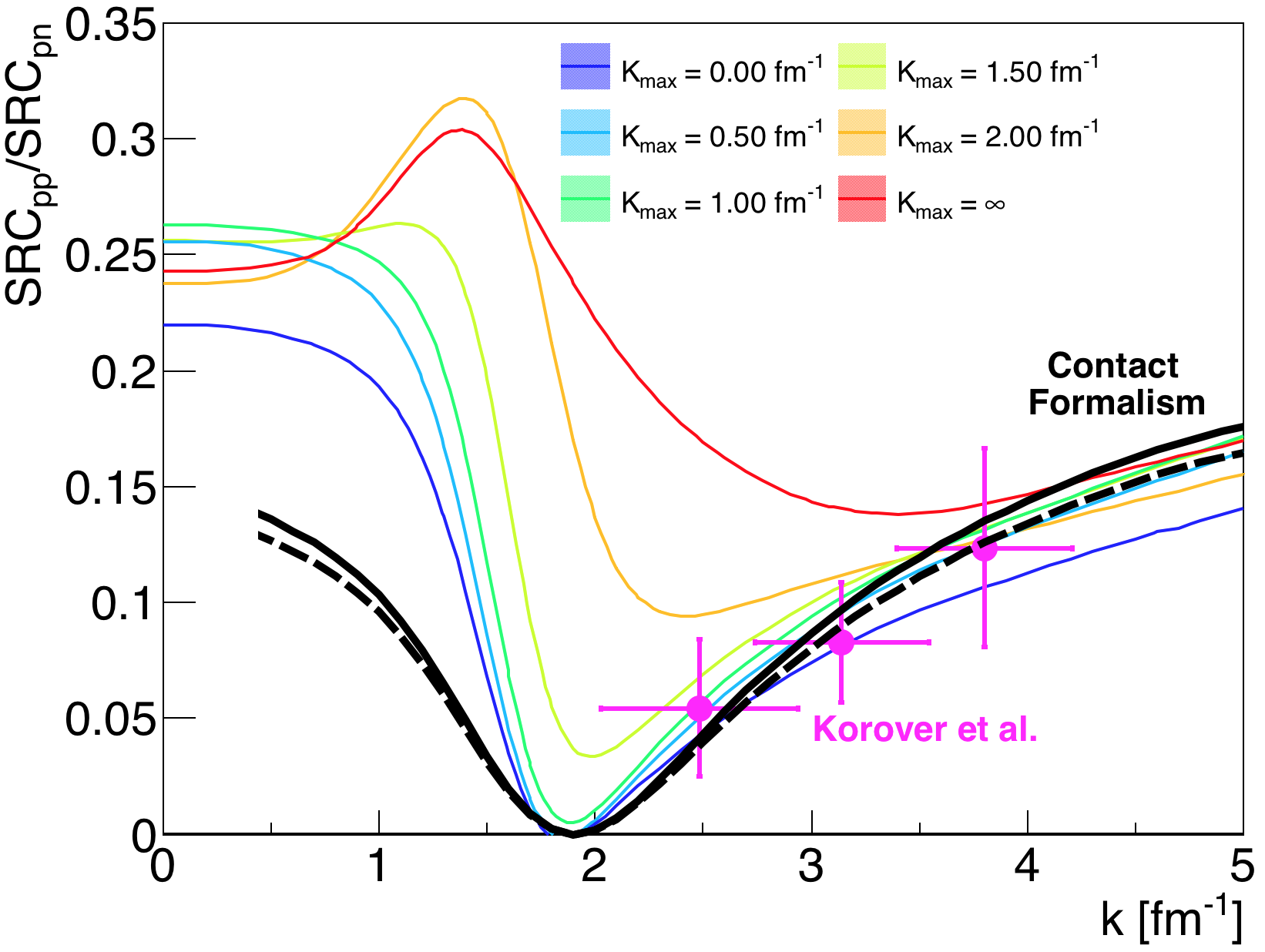}
\caption{\label{fig:pp_np} The ratio of proton-proton to proton-neutron SRC pairs in $^4$He as a function of the pair momentum extracted from $^4$He(e,e'pN) measurements \cite{korover14}. The colored lines show the equivalent ab-initio two-body momentum densities ratio integrated over the c.m. momentum from 0 to $K_{max}$ that varies from zero to infinity \cite{wiringa14}. The solid (dashed) black line is the contact theory prediction calculated using the contacts extracted in momentum (coordinate) space (\eq{eq:exp2}). }
\end{figure}

These two approaches can be demonstrated 
by comparing the two-body density calculations to data. 
Fig.~\ref{fig:pp_np} shows the calculated and measured proton-proton ($pp$)
 to proton-neutron ($pn$) pairs density ratio in $^4$He as a function of their relative momentum.
The experimental data are obtained from recent electron induced two-nucleon knockout measurements
performed in kinematics dominated by breakup of SRC pairs \cite{korover14}.
The calculated pair density ratio is shown as a function of the relative pair momentum
and is given by: $\int_0^{K_{\rm{max}}}d\Kvec F_{pp}(\kvec,\Kvec)/\int_0^{K_{\rm{max}}}d\Kvec F_{np}(\kvec,\Kvec)$,
where $K_{max}$ varies from zero to infinity. 
As can be seen, as long as the maximal c.m momentum is small, 
i.e. $K_{max}<1-1.5$ fm$^{-1} \sim k_F$, the calculated ratio
describes well the experimental data for $k>k_F$. This demonstrates
the above second approach. These results are inline with those of Ref. \cite{Alvioli:2016wwp}.
On the other hand, demonstrating the first approach,
if we concentrate on very high relative momentum,
i.e. $k>4$ fm$^{-1}$, we can see that the ratios are largely insensitive
to the value of $K_{max}$.

Equipped with these observations, we are now in position to utilize
the two-body densities to extract the values of the nuclear contacts.

%


{\it Extracting the nuclear contacts --}
As explained above, we consider  four main nuclear contacts: 
 singlet $\ell=0$ $pn$, $pp$, and $nn$, and  triplet $pn$ deuteron channel. 
For symmetric nuclei, spin-zero $pp$ and $nn$ pairs are identical, leaving three nuclear contacts: 
$C_{nn}^{s=0}$, $ C_{pn}^{s=0}$, and $ C_{pn}^{s=1}$. Isospin symmetry can be used to relate the various $s=0$ contacts, leaving two independent contacts: spin-singlet and spin-triplet. For what follows, we do not impose isospin symmetry in order to study its manifestation in the case of SRC pairs.

We will extract the values of the contacts for nuclei up to $^{40}$Ca in three different methods.
In the first two methods we use the available two-body densities \cite{wiringa14}, in momentum space
and coordinate space, separately. In the third method, we use experimental data.
The results are summarized in Table~\ref{tab:contacts}, where one can see a good agreement
between all three methods.

In the first (second) method, the values are 
extracted by fitting the factorized two-body momentum (coordinate) space expressions
of Eq.~(\ref{eq:2bmd}) to the equivalent two-body density obtained from many-body VMC
calculations \cite{wiringa14}.
The $s=0$ $pp$ and $nn$ contacts are obtained by fitting the VMC $pp$ and $nn$ two-body density respectively. 
The $s=1$ and $s=0$ $pn$ contacts are obtained from simultaneously fitting the spin-isospin $ST=10$ $pn$ two-body density 
and the total $pn$ two-body density. 
In view of the discussion above, the fitting range was 4 fm$^{-1}$ to 4.8 fm$^{-1}$,
as $F_{NN}(\kvec)$ is dominated by SRC pairs only for $k > 4 \;\rm{fm}^{-1}$. In coordinate space the fitting was done in the range from 0.25 fm to 1 fm.
The determination of the uncertainties is described in \cite{Suplementary}.
As VMC coordinate space distributions are not available for the different spin-isospin states,
we assumed isospin symmetry (i.e. equal $s=0$ contacts) for the symmetric nuclei.
The universal functions $\varphi^\alpha_{ij}$ were calculated using the AV18 potential
\cite{Suplementary}.

Experimentally, the nuclear contacts can be evaluated using the measured $pp$-to-$pn$ SRC
 pairs ratio discussed above, $\frac{SRC_{pp}}{SRC_{pn} }(k)$, and the high-momentum scaling factor, $a_2(A/d)$. 
The latter is extracted from large momentum transfer inclusive electron scattering cross-section ratios and determines the 
relative number of high-momentum ($ k > k_F$) nucleons in a nucleus, $A$, 
relative to deuterium \cite{frankfurt93, egiyan02, egiyan06, fomin12, Hen12}, 
assuming the effects of Final-State Interactions and other reaction channels are suppressed in the kinematics of these measurements due to the large momentum transfer and the use of cross-section ratios, see Ref.~\cite{Hen:2016kwk,Atti:2015eda} and references therein for details. 
Within the contact formalism, these experimental quantities can be expressed as:
\begin{equation}
  a_2(A/d) \int_{k_{F}}^{\infty}|\tilde{\psi}_{d}(\kvec)|^2 d\kvec = 
              \frac{C_{nn}^{s=0}+C_{pp}^{s=0}+C_{pn}^{s=0}+C_{pn}^{s=1}}{A/2}
\label{eq:exp1}
\end{equation}
\begin{equation}
  \frac{SRC_{pp}}{SRC_{pn}}(k) = 
        \frac{C_{pp}^{s=0}|\tilde{\varphi}^{s=0}_{pp} (k)|^2 }
             {C_{pn}^{s=0}|\tilde{\varphi}^{s=0}_{pn} (k)|^2+C_{pn}^{s=1}|\tilde{\varphi}^{s=1}_{pn}(k)|^2}
\label{eq:exp2}
\end{equation}
where $\tilde{\psi}_{d}(\kvec)$ is the deuteron wave function, normalized to one.
In Eq. \eqref{eq:exp2} it is assumed that
the c.m. motion of SRC pairs is small, 
and similar for the different types of pairs in a given nucleus, as observed experimentally
\cite{subedi08, korover14, hen14, shneor07}.
The experimental values of the contacts, shown in Table~\ref{tab:contacts},
were extracted for symmetric nuclei using these relations, assuming isospin symmetry. 


\begin{table*}[t]
  \renewcommand{\arraystretch}{1.2}
  \begin{tabular}{|c|c|c|c|c|c|c|c|c|}
    \hline\hline
    \multirow{2}{*}{\textbf{A}} & \multicolumn{4}{c|}{\textbf{k-space}} & \multicolumn{4}{c|}{\textbf{r-space}}\\
    \cline{2-9}
    & $C_{pn}^{s=1}$ & {$C_{pn}^{s=0}$} & {$C_{nn}^{s=0}$} & {$C_{pp}^{s=0}$} & {$C_{pn}^{s=1}$} & {$C_{pn}^{s=0}$} & {$C_{nn}^{s=0}$} &      
      {$C_{pp}^{s=0}$}\\
    \hline\hline
    \multirow{2}{*}{\textbf{$^4$He}} & 12.3$\pm$0.1 & 0.69$\pm$0.03 & \multicolumn{2}{c|}{0.65$\pm$0.03} & \multirow{2}{*}{11.61$\pm$0.03} & \multicolumn{3}{c|}{\multirow{2}{*}{0.567$\pm$0.004}}\\
    \cline{2-5}
    &14.9$\pm$0.7 (exp) & \multicolumn{3}{c|}{0.8$\pm$0.2 (exp)} & & \multicolumn{3}{c|}{} \\
    \hline
    \textbf{$^6$Li} & 10.5$\pm$0.1 & 0.53$\pm$0.05 & \multicolumn{2}{c|}{0.49$\pm$0.03} & 10.14$\pm$0.04 & \multicolumn{3}{c|}{0.415$\pm$0.004}\\
    \hline
    \textbf{$^7$Li} & 10.6 $\pm$ 0.1 & 0.71 $\pm$ 0.06 & 0.78 $\pm$ 0.04 & 0.44 $\pm$ 0.03 & 9.0 $\pm$ 2.0 & 0.6 $\pm$ 0.4 & 0.647 $\pm$ 0.004 & 0.350 $\pm$ 0.004 \\
    \hline
    \textbf{$^8$Be} & 13.2$\pm$0.2 & 0.86$\pm$0.09 & \multicolumn{2}{c|}{0.79$\pm$0.07} & 12.0$\pm$0.1 & \multicolumn{3}{c|}{0.603$\pm$0.003}\\
    \hline
    \textbf{$^9$Be} & 12.3$\pm$0.2 & 0.90$\pm$0.10 & 0.84$\pm$0.07 & 0.69$\pm$0.06 & 10.0$\pm$3.0 & 0.7$\pm$0.7 & 0.65$\pm$0.02 & 0.524$\pm$0.005 \\
    \hline
    \textbf{$^{10}$B} & 11.7$\pm$0.2 & 0.89$\pm$0.09 & \multicolumn{2}{c|}{0.79$\pm$0.06} & 10.7$\pm$0.2 & \multicolumn{3}{c|}{0.57$\pm$0.02}\\
    \hline  
    \multirow{2}{*}{\textbf{$^{12}$C}} & 16.8$\pm$0.8 & 1.4$\pm$0.2 & \multicolumn{2}{c|}{1.3$\pm$0.2} & \multirow{2}{*}{14.9$\pm$0.1} & \multicolumn{3}{c|}{\multirow{2}{*}{0.83$\pm$0.01}}\\
    \cline{2-5}
    &18$\pm$2 (exp)& \multicolumn{3}{c|}{1.5$\pm$0.5 (exp)} & & \multicolumn{3}{c|}{} \\
    \hline
    \textbf{$^{16}$O} &      \multicolumn{4}{c|}{ } & 11.4$\pm$0.3 & \multicolumn{3}{c|}{0.68$\pm$0.03}\\
    \hline
    \textbf{$^{40}$Ca} &       \multicolumn{4}{c|}{ } & 11.6$\pm$0.3 & \multicolumn{3}{c|}{0.73$\pm$0.04}\\
    \hline\hline
  \end{tabular}
  \caption{The nuclear contacts for a variety of nuclei. The contacts are extracted by
fitting the asymptotic expressions of Eq. ~(\ref{eq:2bmd}) to the VMC two-body densities in momentum (k) and coordinate (r) space separately. For $^4$He and $^{12}$C the contacts extracted from electron scattering data are also shown. The nuclear contacts are divided by A/2 and multiplied by 100 to give the percent of nucleons above $k_F$ in the different SRC channels.}
  \label{tab:contacts}
\end{table*}

The agreement between the values of the contacts that were extracted in momentum and
coordinate space, points to a quantitative equivalence between high-momentum and short-range
physics in nuclear systems. The agreement with the experimental extraction is an
important indication for the validity of the contact formalism to nuclear systems.
Another interesting feature of the extracted values is that, for symmetric nuclei,
the momentum space $s=0$ $pp$ and $pn$ contacts are the same within uncertainties, 
in contrast to combinatorial expectations. 

\begin{figure} [b]
\includegraphics[width=0.5\textwidth]{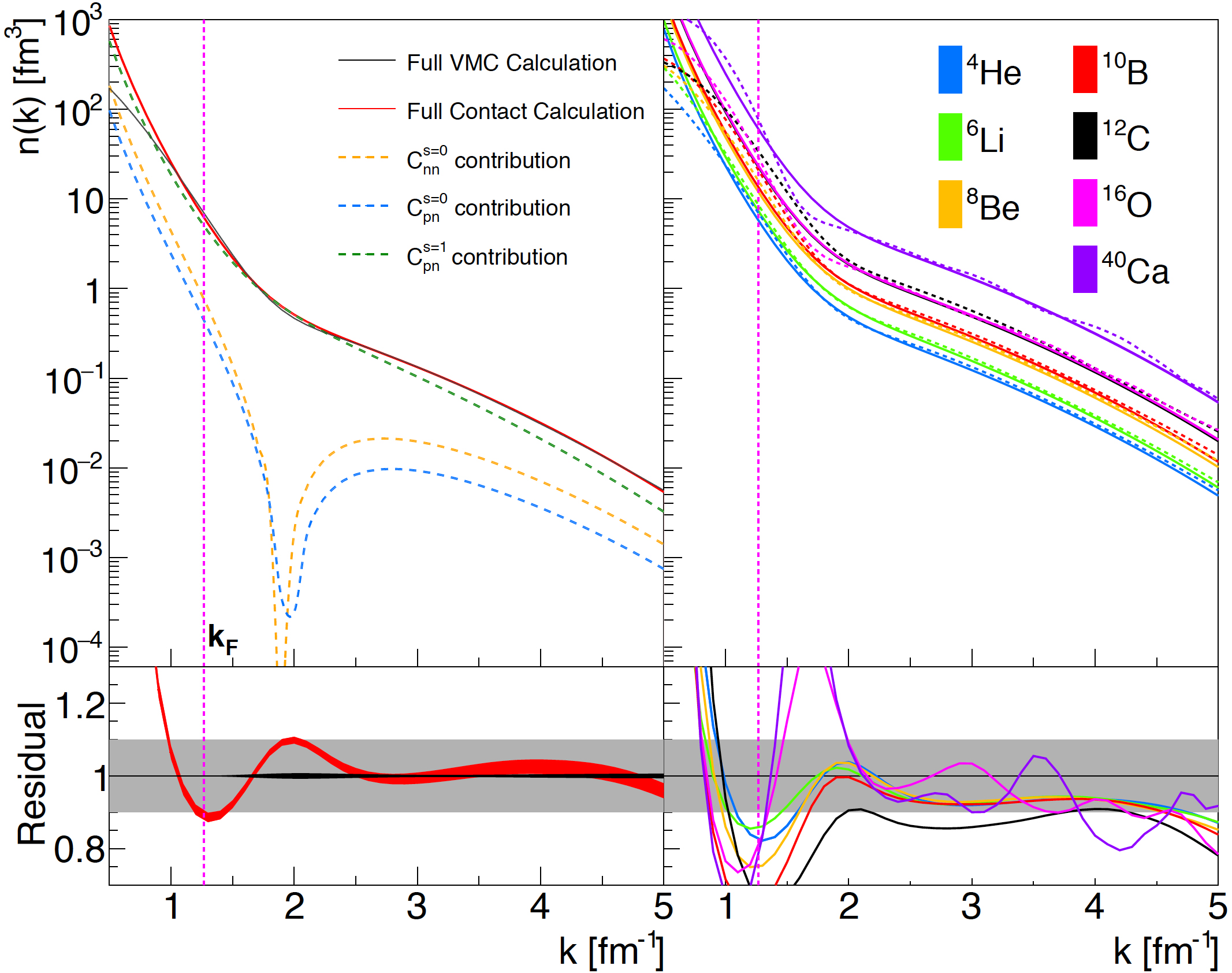}
\caption{\label{fig:1bmdist} (left) $^4$He one-body momentum densities extracted from ab-initio VMC calculations (solid black band) and using the nuclear contact formalism (solid red band). The dashed lines show the contribution of different channels to the total contact calculation, 
using the contacts extracted in momentum space. 
The residual plot shows the ratio of the contact calculations to the VMC. The shaded region marks the 10$\%$ agreement region. The width of the black and red lines represents the individual uncertainties 
in the calculations. 
(right) The same, without error bands, comparing VMC calculations (dashed lines) and
the nuclear contact formalism (solids lines) for different nuclei. The contacts
used to calculate the distributions on the right plot were extracted in coordinate space.
}
\end{figure}

We can now utilize the values of the contacts to further
investigate the predictions of the theory.
First, we note that as the relation between the contacts and the
one body momentum distribution, given in Eq. \eqref{eq:1bmdist},
was not used to fit the values of the contacts it can
be considered as a verifiable prediction.
Fig.~\ref{fig:1bmdist} compares, for several nuclei, the one-body momentum distribution obtained 
from many-body VMC calculations to the prediction of 
Eq.~(\ref{eq:1bmdist}). 
As can be seen, the asymptotic 1-body density, as predicted by 
the contact theory, reproduces with 10$\%$-20$\%$  accuracy the many-body calculation starting from 
$k_F$ to 5 fm$^{-1}$, where the momentum density varies over 3 orders of magnitude.
It is worth emphasizing that even though the contacts fitting range was only
$k>4$ fm$^{-1}$ using the two-body momentum distribution, the one-body
momentum distribution is reproduced starting from $k_F$, as expected.

The contacts can also be used to calculate the $pp$ to $pn$ SRC pairs ratio
using Eq. \eqref{eq:exp2}. This ratio can
be compared with experimental electron induced two-nucleon knockout data \cite{tang03,piasetzky06, subedi08, korover14, hen14} as shown for $^4$He in Fig.~\ref{fig:pp_np}. 
A similar comparison for $^{12}$C \cite{subedi08} also shows a good agreement~\cite{Suplementary}.
We can see that the contact predictions are in a good agreement
with the experimental results and ab-initio calcualtions.

The contact formalism also allows us to evaluate the contributions
of the different two-body channels to SRC pairs. Such decomposition is shown
in Fig. \ref{fig:1bmdist} (left panel) for $^4$He.
The values of the contacts clearly show the expected
dominance of the deuteron channel in SRC pairs.
The fact that the contact formalism reproduces the VMC one-body momentum density 
 to 10$\%$ - 20$\%$ accuracy, without utilizing the spin-isospin $ST=11$ channels, indicates their small importance
to SRCs in the nuclei considered here. This stands in contrast to other works that do find a non-negligible
contribution of $ST=11$ pairs \cite{Feldmeier:2011qy, Alvioli:2012qa}. 
A possible explanation for this difference goes back to our discussion of
the regions where the two-body momentum distribution describes SRCs.
In these two papers, the c.m. momentum was not limited to small values and, thus, contributions from non-correlated pairs are expected to be significant.
The contact theory provides a simple framework to perform such decompositions
for SRC channels.



{\it Conclusions --}
Even though nuclear systems do not strictly fulfill the scale-separation 
conditions required by the contact theory, both ab-initio one body momentum 
distribution above $k_F$ and the experimental data are well reproduced using
factorized asymptotic wave-functions and nuclear contact theory.

Consistent contacts extracted by separately fitting coordinate and
momentum space two-body densities show equivalence between
high-momentum and short-range dynamics in nuclear systems. 
Experimental extraction of the contacts gives also similar results.
The values of the contacts allow a proper analysis of the spin-isospin
quantum numbers of SRC pairs, and also 
reveal the non-combinatorial isospin-spin symmetry of SRCs.

This work provides clear evidence for the applicability of
the generalized contact formalism to nuclear systems,
and open the path towards further SRC studies.


\begin{acknowledgments}
We would like to thank B. Bazak, W. Cosyn, C. Ciofi degli Atti, S. Gandolfi, G. Miller, E. Pazi, 
J. Ryckebush, M. Sargsian, M. Strikman, and L.B. Weinstein for many fruitful discussions. 
This work was supported by the Pazy foundation, the Israel Science Foundation (grant no. 136/12, and 1334/16),
and the U.S. Department of Energy Office of Science, Office of nuclear physics program under award number
DE-FG02-94ER40818.
\end{acknowledgments}

\bibliography{Contact_bib}

\end{document}